\documentclass[11pt,twoside]{article}

\usepackage{asp2006}
\usepackage{epsf}
\usepackage{psfig}
\usepackage{lscape}

\begin{document}
\title{Planets of young stars}   
\author{Eike W. Guenther and Massimiliano Esposito}   
\affil{Th\"uringer Landessternwarte Tautenburg, 07778 Tautenburg, Germany}    

\begin{abstract}

Since the first massive planet in a short period orbit was discovered,
the question arised how such an object could have formed. There are
basically two formation scenarios: migration due to planet-disk or
planet-planet interaction. Which of the two scenarios is more
realistic can be found out by observing short-period planets of stars
with an age between $10^7$ and $10^8$ yrs. The second aim of the
survey is to find out how many planets originally formed, and how many
of these are destroyed in the first Gyrs: Do most young, close-in
planets evaporate, or spiral into the host stars? In here we report
on the first results of a radial-velocity search program for planets
of young stars which we began in 2004. Using HARPS, we currently
monitor 85 stars with ages between $10^7$ and $10^8$ yrs. We show that
the detection of planets of young stars is possible. Up to now, we
have identified 3 planet-candidates. Taking this result together with
the results of other surveys, we conclude that the frequency of
massive-short period planets of young stars is not dramatically higher
than that of old stars.

\end{abstract}


\section{The idea for the survey}  

In the last 11 years, about 200 extrasolar planets have been detected
indirectly by means of precise radial velocity (RV) surveys. The
detection of now 16 transiting planets, as well as astrometric
observations of $\epsilon$ Eri and the detection of the thermal emission
of $\mu$ Andromeda b with the Spitzer satellite demonstrates that the
vast majority of the objects found by RV-surveys are in fact planets
(Charbonneau et al. (2000); Harrington et al. (2006); Benedict et
al. (2006)). The orbits of the extrasolar planets are surprisingly
different from those in our solar system, as most of the long-period
planets have rather eccentric orbits, and as there are planets with very
short orbital periods (so called ``hot Jupiters'', or ``Pegasides''
after the first such object found.). Up to now 49 extrasolar planets
have been found which have a semi-major axis of $\leq 0.1$ AU (orbital
periods $\leq 10$ days).  When corrected for selection effects, the true
frequency of hot Jupiters orbiting old stars is few \% (Lineweaver \&
Grether, \cite{lineweaver03}).

The discovery of hot Jupiters was very surprising, and their formation
still poses a mystery. In the standard model massive planets form via
core accretion: In the first step a solid core of about 5 to 10
$M_{earth}$ forms, which subsequently accretes gas from the disk in
order to form a $M_{jupiter}$-planet (see for example the review on
planet formation by Perryman (2000)). The core accretion model is
supported by the discovery of the eclipsing planet HD149026 b which
has a $\sim$ 67 $M_{earth}$ core consisting of heavy elements (Sato et
al.  2005), and by the fact that stars with an overabundance of heavy
elements also have a higher frequency of massive planets (Santos et
al. 2004). In-situ formation of 'hot Jupiters' is unlikely, as it
requires a large surface density of dust particles in the inner
disk. Such large surface density of dust particles is considered
unlikely, because the temperature in the inner 0.1 AU of the disk is
above the sublimation temperature of the dust.  Interferometric
observations of young stars in fact show that there is basically no
dust within 0.1 AU of the star (Akeson et al.  2005).  With in-situ
formation being unlikely, there are only two scenarios for the
formation of close-in planets:

\noindent
{\bf 1.) Planet migration via interaction with the disk:} Giant
planets in a circum\-stellar disk can migrate inwards by torques
between the planet and the disk. The migration-rate due to interaction
with the disk is over large range independent of the mass of the
planet but depends linearly on the mass of the disk (Lufkin et
al. 2004). In general migration will damp the eccentricities so that
even young close-in planets should have round orbits. In this scenario
7\% of the stars should have planets of more than 0.6 $M_{jupiter}$
within 0.1 AU at the end of the formation period (age $10^7$ to $10^8$
yrs; Armitage et al.  2002). However, as shown by D'Angelo et
al. (2006) this picture changes dramatically, if an eccentric disk is
considered.  A planet in an eccentric disk will also have an eccentric
orbit, and the eccentricity of the planet will even increase with
time.  If the eccentricity becomes larger than 0.2 the planet will
migrate outward instead of inward.

\noindent
B{\bf 2.) Migration due to planet-planet interaction:} If a system of
three or more giant planets forms, their orbits are stable during
their formation epoch ($\leq 10^7$ yrs).  But, after the depletion of
the disk gas, mutual gravitational perturbation between the planets
will induce a gradual increase in their orbital eccentricities, until
their orbits become unstable. Subsequent gravitational encounters
among these planets can lead to the ejection of the planets from the
system while placing others into highly eccentric orbits, both closer
and farther from the star (Weidenschilling \& Marzari 1996). In this
scenario, the close-in planets form at an age of about $10^7$ yrs, and
the orbits will initially be eccentric. After $10^9$ yrs the orbits
of short-period planets will be round due to tidal interaction.

In order to find out which of the two scenarios is more realistic, we
have to observe stars with ages between $10^7$ to $10^8$ yrs.  In the
case of the first scenario about 7\% of the stars will have a hot
Jupiter with a round orbit. In the case of the second scenario, the
interaction between planets would happen at an age between $10^7$ to
$10^8$ yrs. If the second model were correct, we may find planets in
instable orbits.  At least we should find close-in planets with
eccentric orbits.  

Another interesting aspect of our project is that we can find out how
many planets originally formed, and how many of them are destroyed
within the first Gyrs.  There are several possibilities how planets can
be destroyed even after the proto\-planetary disks have been dispersed.
Close-in extrasolar planets experience strong tidal interactions with
their central star. If the planet's orbital period is shorter than the
star's rotation period, tidal friction will lead to a spin-up of the
star and, thus also lead to a decrease of the semi-major axis of the
planet's orbit (P{\"a}tzold et al.  2004). It is thus in principle
possible that planets may even spiral into the host star. Close-in
planets may also evaporate due to the XUV radiation of the star. It has
been estimated that planets with an orbital distance $\leq$ 10 days
($\sim$ 0.1 AU) loose between 20\,
first Gyrs, were the XUV flux of the host star is particularly strong
(Baraffe et al.  2006).  If this is true, all hot Neptunes orbiting old
stars would have been short period Jupiters at young age, and we should
find many massive, close-in planets in our survey.

\section{The survey}  

For our survey, we selected 92 stars in the TWA Hydra [10-30 Myr],
$\beta$ Pic [$\sim$ 12 Myr], Horologium [10-30 Myr], Tucana
[$\sim$ 40 Myr], IC2391 [30-40 Myr] association and a few stars of
similar age in the field. We were surprised to find out that 7 of the
stars selected from the literature turned out to be old, as they have no
LiI-absorption, and the RV-variations are $\leq $ 5 $m\,s^{-1}$.

The survey is being carried out with HARPS on the ESO 3.6-m-telescope
at La Silla, started in the 2004 and is still ongoing. Up to now, we
have taken 558 spectra of the 85 young stars. We use the
ThAr-simultaneous reference method and the RV-pipeline developed by
the Geneva group (Mayor et al. 2003). All observations were carried
out in service mode and spectra were taken at random time during the
semester.

\section{The influence of stellar activity}  

Of course young stars are less ideal for RV-planet-search-projects,
because they are active. The question thus is, whether it is possible
to detect close-in massive planets, or not.  Previous to this survey,
we have carried out RV-measurements of classical and weak-line T Tauri
stars. We found that 50\% of the stars show RV-variation of $\leq$ 750
$m\,s^{-1}$, and that the RV-jitter can be as large as 2 or 3
$km\,s^{-1}$ (Fig.\,1).  This limits the possibility to detect
companions with orbital periods of 10 days to masses that are in the
brown dwarf regime. Thus, the detection of planets of T Tauri stars by
means of precise RV-measurements is extremely difficult, if not
impossible. In the case of the stars with an age between $10^7$ to
$10^8$ yrs, we find that more than 50\% of the stars show
RV-variations of $\leq$ 100 $m\,s^{-1}$. This allows us to detect
planets with the mass of Jupiter with orbital periods of $\leq$ 10
days.

\begin{figure}[!ht]
\vspace*{7.0cm}
\begin{center}
\includegraphics{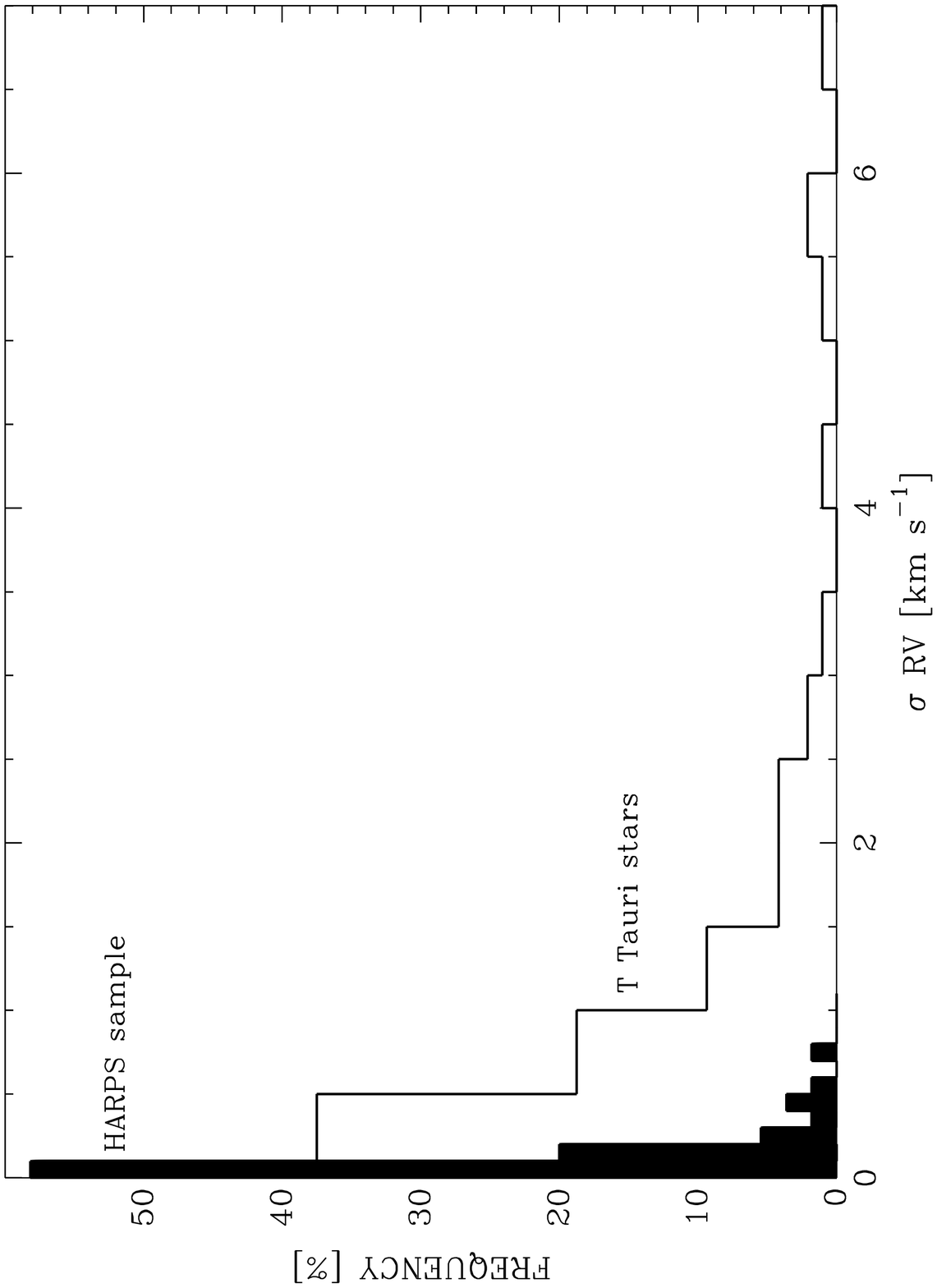}
\caption{RV-variations of T Tauri stars and the stars of this
survey. While T Tauri stars are so active that planet detection via
RV-measurements is not possible, the stars of this survey are
sufficiently inactive to allow the detection of massive planets with short
orbital periods.}
\end{center}
\end{figure}

For detecting planets of active stars, it is necessary to carry out a
number of tests in order to demonstrate that the RV-variations are
caused by an orbiting planet and not by stellar activity. In
principle, the spectral-lines of a spotted star are red-shifted, when
the spot rotates into view, and blue-shifted when the spot rotates
out of sight.

\begin{itemize}
\item A spot on the stellar surface will cause a hump in the line-profile.
      In a spotted star the line-asymmetry will thus be correlated with the RV.
\item Spots on the stellar surface will cause brightness variations. In a
      spotted star the brightness variations are thus related to the RV-variations. 
\item Plage regions on the sun appear bright in images taken in the emission
      core of the Ca\,II\,H and K-lines. Thus, the presence and absence
      of active regions on the stellar surface can be inferred from the 
      strength of the emission cores of these lines. In a spotted star, the
      emission cores are thus correlated with the stellar activity. Other
      chromospheric lines like H$\alpha$ may also be used for similar purposes.
\item Spots have a limited life time. The RV-signal of spotted star thus
      changes with time. However, we should keep in mind that there are some stars
      were the spot-pattern remained unchanged for years.
\item Since the difference in brightness between a spot and the photosphere is
      much smaller in the infrared than in the optical regime, the RV-variations
      of a spotted star is much smaller at infrared wavelength than in the 
      optical. 
\end{itemize}

\section{Binaries}  

The first companions found were stellar companions. In the following, we
will give a short overview of the binaries found.

\noindent {\bf HD113449:} We have taken 14 RV-measurements of this star
which is a member of the AB Dor moving group (Zuckerman, 2004) and found
that it to be a binary with an orbital period of 221 days. The
m\,sin\i\, of the companion is about half a solar-mass.

\noindent
{\bf HD35850:} Although we have taken only 4 spectra of this star, the
large RV-variations indicate that it must be a binary, possibly of short
period. This star is member of the $\beta$ Pic moving group (Zuckerman
\& Song, 2004).

\noindent
{\bf V343Nor:} We have taken 14 RV-measurements of this star, and
found a linear trend of $1.7\pm0.1$ $km\,s^{-1}\,year^{-1}$. It thus
must be a binary with a long orbital period.  V343Nor is again a
member of the $\beta$ Pic moving group (Zuckerman \& Song 2004).

\noindent
{\bf HD13183:} Although we have taken only 5 measurements up to now,
we conclude that also this star must be a binary with a long orbital
period.  HD13183 is a member of the Tucana/Horologium Association
(Zuckerman \& Song 2004).

\noindent 
{\bf HD181321:} We have taken 10 RV-measurements of this star, and found
a linear trend of $1.4\pm0.1$ $km\,s^{-1}\,year^{-1}$, which is possibly
slightly curved. This star was identified by Wichmann et al. (2003) as a
young star.

\section{A very interesting candidate}  

Amongst the 85 young stars surveyed, we found three stars which show
RV-variations with amplitudes between 200 and 400 $m\,s^{-1}$ and
periods of a few days. One of these is particularly interesting: We
observed this for 1.3 years, and found only one very clean period of
5.02 days (Fig.\,2). The RV-values also phase up very nicely with
this period (Fig.\,3).  Thus, the RV-variations must either be caused
by a long-lived active region, or by an orbiting object with an
m\,sin\,i of 0.8 $M_{Jupiter}$. Active regions of stars usually change
within a year, and thus the periodogram of active stars usually does
not look as clean as Fig.\,2 (see K\"onig et al. (2005) for details).
However, there are stars were the spot-pattern did not change in
years. For example in the case of V410 Tau, the spot pattern remained
unchanged for 10 years (Fern\'andez et al. 2004; Stelzer et al. 2003)!

In order to find out whether the RV-variations are caused by stellar
activity, or by an orbiting planet, we thus have to undertake all the
tests mentioned above. The outcome of the first test is shown in
Fig.\,4. The figure shows the normalised equivalent width of the CaH,
CaK, H$\alpha$, H$\beta$, and H$\gamma$ emission lines phased to the
5.02 day period. The equivalent widths of the lines certainly do not
phase up with this period.  We are currently carrying out photometric
observations and measuring the line-asymmetry. More work still needs
to be done until a definite conclusion can be drawn but up to now this
candidate looks promising. The best, and most critical test will be
to carry out RV-measurements at infrared wavelengths.

\begin{figure}[!ht]
\vspace*{7.0cm}
\begin{center}
\includegraphics{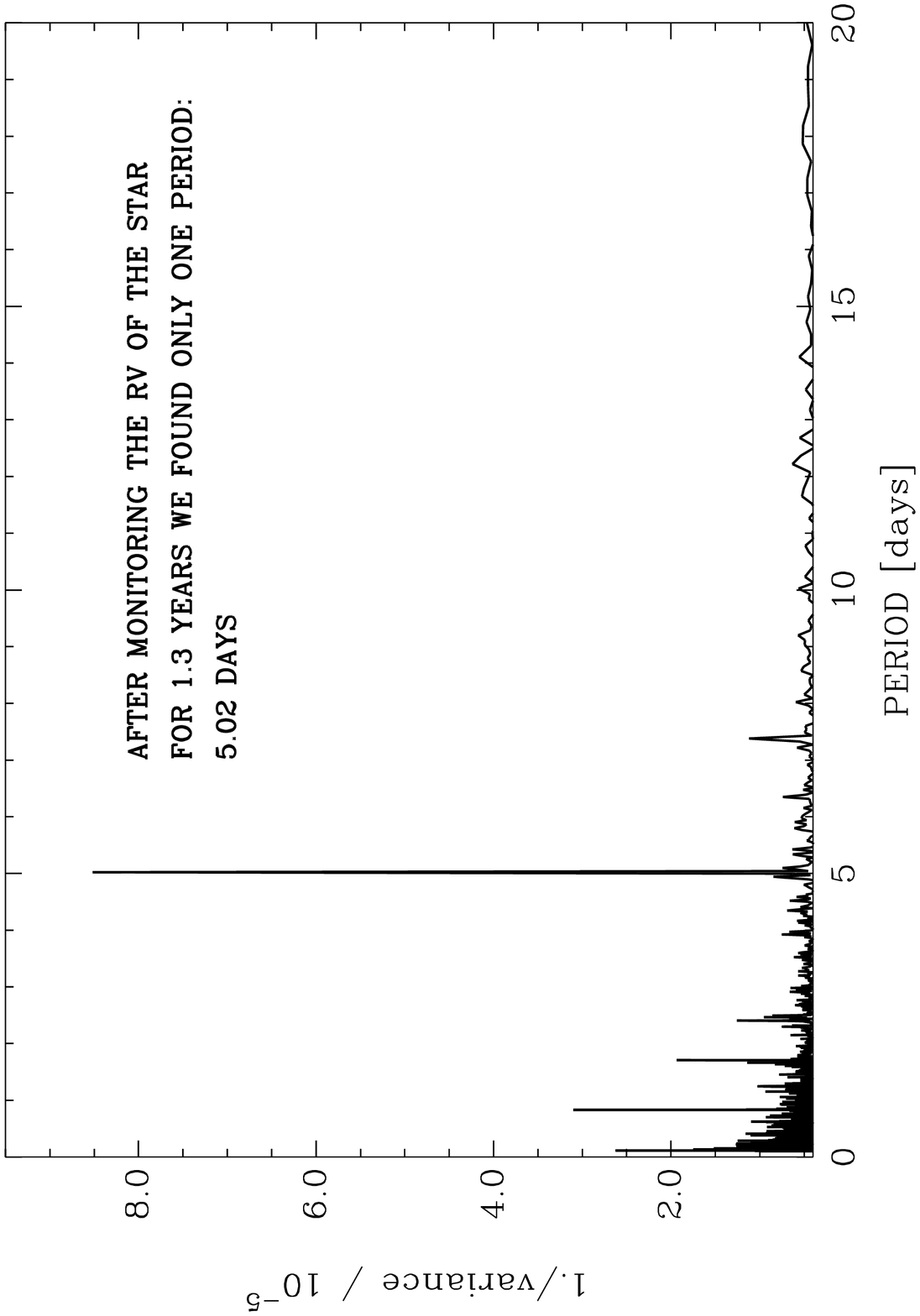}
\caption{Periodogram of the RV-variation of our best candidate. There is
only one significant period.}
\end{center}
\end{figure}

\begin{figure}[!ht]
\vspace*{7.0cm}
\begin{center}
\includegraphics{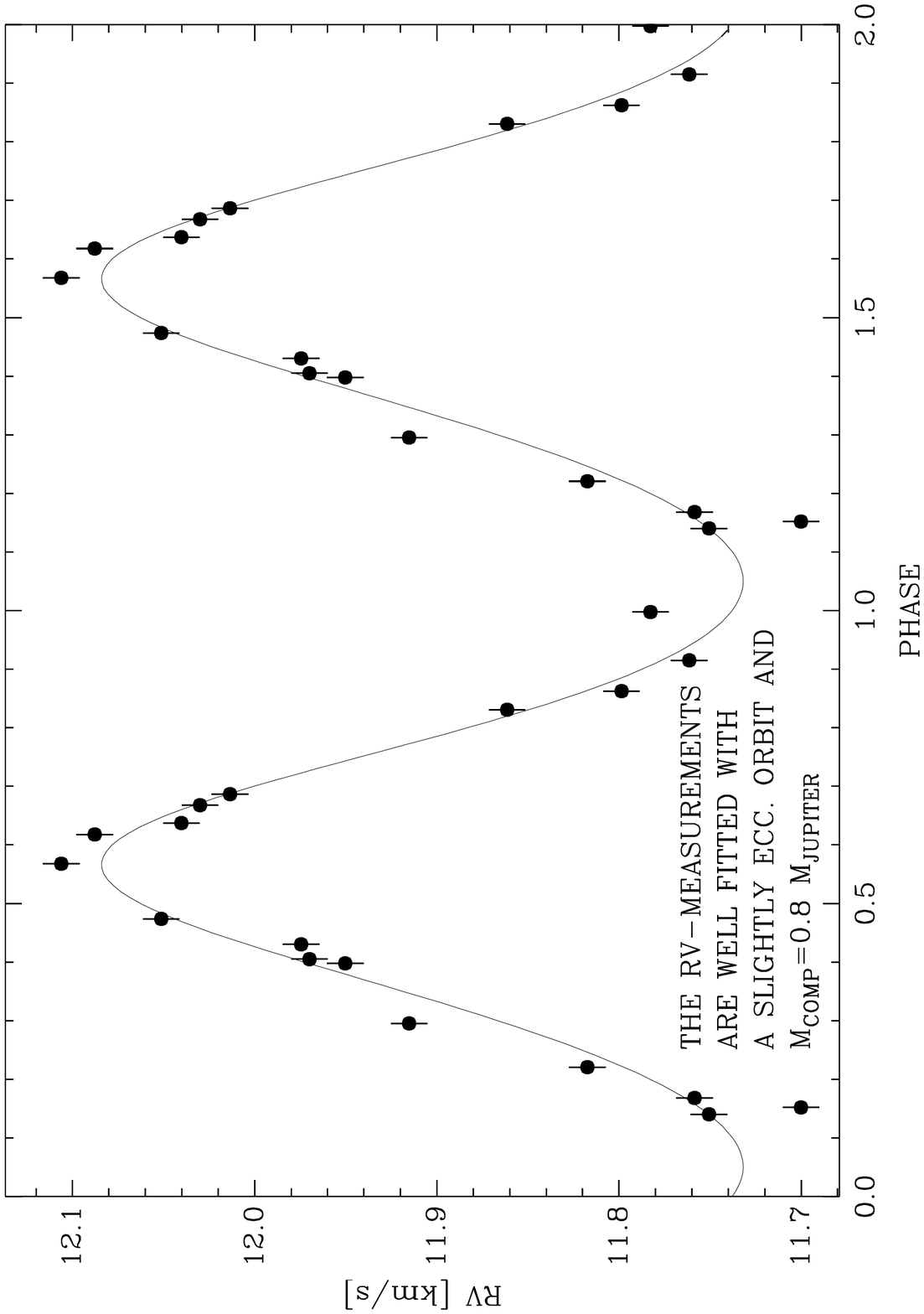}
\caption{RV-variations phased to a period of 5.02 days.
Since we monitored the star for 1.3 years, the periodic RV-variations
must either be caused by a very stable active region, or by
an orbiting planet.
}
\end{center}
\end{figure}

\begin{figure}[!ht]
\vspace*{7.0cm}
\begin{center}
\includegraphics{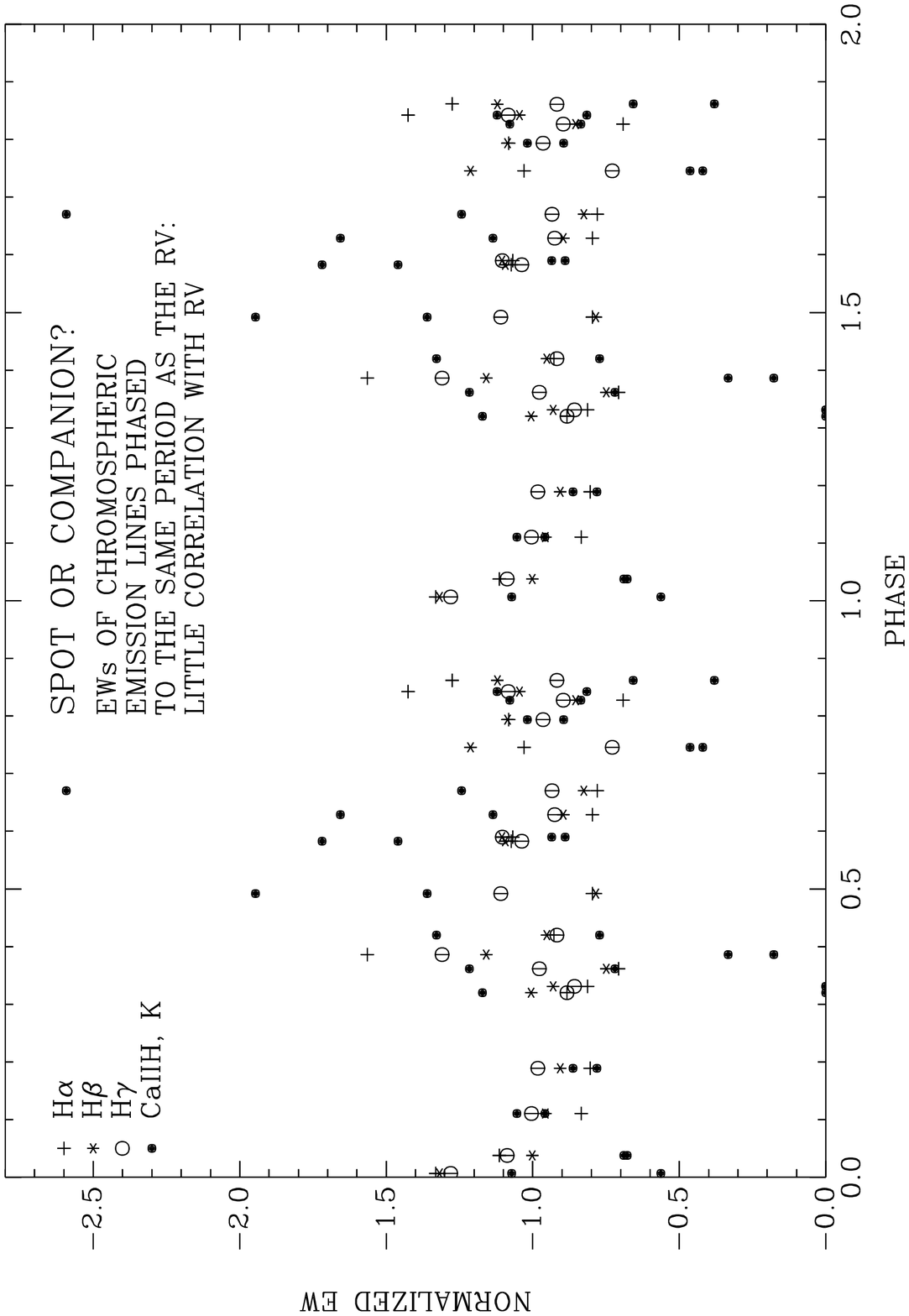}
\caption{In order to find out whether the RV-variations are caused by
stellar activity, or by an orbiting planet, we determined the
equivalent width of all the chromospheric lines.  The figure shows the
normalised equivalent widths of the CaH, CaK, H$\alpha$, H$\beta$, and
H$\gamma$ emission lines phased to the same period as above. From the lack 
of correlation, we conclude that the RV-variation are unlikely to be caused
by stellar activity.}
\end{center}
\end{figure}

\section{Putting things in perspective}  

Additionally to this survey we have also carried out a similar survey in
the northern hemisphere using the 2-m-Alfred Jensch telescope in
Tautenburg. In that survey we observed 46 young stars (Esposito et
al. 2006). Most of the stars in that observed have ages comparable to
the Pleiades ($\sim100$ Myr) but some are $300 \div 500$ Myr old. In
that survey we found only one good candidate. The period of the
RV-variations of this object is only 1.3 days, and the amplitude 51
$m\,s^{-1}$. For 19 stars we can exclude planets with an $m\,sin\,i$
$\geq$ 1 $M_{Jupiter}$ and with periods $\leq 10$ days.  For 8
additional stars it is possible exclude planets with an $m\,sin\,i$
$\geq$ 5 $M_{Jupiter}$, and periods $\leq 10$ days.  Thus, in total, we
have served 131 young stars, and found only 4 good planet candidates. Up
to now, we can not say whether any of these is really a planet but we
can already conclude that there is certainly not a large population of
close-in massive planets. This implies that the fraction of close-in
planets that are destroyed, or evaporate in the first $10^7$ to $10^9$
yrs is low. This result is in good agreement with other studies:

Paulson \& Yelda (2006) observed the $\beta$ Pic ($\sim$ 12 Myr)
association, IC 2391 (30-40 Myr), the Castor moving group, and the Ursa
Major association ($\sim$ 300 Myr), as well as stars of similar age in
the field, 61 stars in total.  They can rule out companions of these
stars with a mass of more than one $M_{Jupiter}$ and an orbital period
shorter than 6 days.  Paulson et al. (2004) also monitored 98 stars in
the Hyades (age $\sim$ 700 Myr). They found four stars with periodic
RV-variations but after monitoring these stars photometrically, they
could rule out in three of the four stars that the RV-variations are
caused by orbiting planets.

Taking everything together, we can thus conclude that the frequency of
massive-short period planets of young stars is not dramatically higher
than that of old stars. It thus seems unlikely to us that the vast
majority of the massive, short-period planets are destroyed in the
first $10^7$ to $10^9$ yrs.


\acknowledgements 

We are grateful to the user support group of ESO/La Silla.  This
research has made use of the SIMBAD database, operated at CDS,
Strasbourg, France.


\end{document}